\documentclass[aps,pre,showpacs,twocolumn]{revtex4}%
\usepackage[english]{babel}
\usepackage{amsmath}
\usepackage{color}
\usepackage{epsfig}
\usepackage{graphicx}
\usepackage{bm}
\usepackage{marvosym}
\usepackage{times,amsmath,amssymb}
\begin{document}
\title{Optimal transport and von Neumann entropy in an Heisenberg XXZ chain out of equilibrium}
\author{Mario Salerno$^{1}$ and Vladislav Popkov$^{2}$}
\affiliation{$^{1}$ Dipartimento di Fisica ``E.R. Caianiello'', Universit\`a di Salerno,
via ponte don Melillo, 84084 Fisciano (SA), Italy}
\affiliation{$^{2}$ Dipartimento di Fisica, Universit\`a di Firenze, via Sansone 1, 50019
Sesto Fiorentino (FI), Italy}
\begin{abstract}
In this paper we investigate the spin currents and the von Neumann entropy (VNE) of an  Heisenberg XXZ chain in contact with twisted XY-boundary magnetic reservoirs by means of the Lindblad master equation. Exact solutions for the stationary reduced density matrix are explicitly constructed for chains of small sizes by using a quantum symmetry operation of the system. These solutions  are then used to investigate the optimal transport in the chain  in terms of the VNE. As a result we show that the maximal spin current always occurs  in the proximity of extrema of the VNE and for particular choices  of parameters (coupling with reservoirs and anisotropy) it can exactly coincide with them. In  the limit of strong coupling we show that minima of the VNE tend to zero, meaning that the maximal transport is achieved in this case with states that are very close to pure states.
\end{abstract}

\date{\today }
\maketitle

\section{Introduction}
There is presently an increasing amount of interest for the understanding of the transport properties in quantum  open (out of equilibrium) systems \cite{Petruccione, PlenioJumps} due to  possible application in the field of condensed matter physics and of quantum computing. In particular, transport properties in anisotropic XXZ Heisenberg  spin chains in contact with magnetic reservoirs have been extensively investigated in the past years  (for extended reviews see Ref. \cite{review1, zotos}) with a range of alternate methods, including exact diagonalizations \cite{zotos96, andrei98},
Bethe-ansatz \cite{shastry, klumper, zotos99}, Lagrange multipliers \cite{Anta97,Anta98}, Lanczos method \cite{lanczos}, quantum Monte Carlo \cite{alvarez}, etc. On the other hand, it is well known that these systems can be also treated by replacing observables associated with the reservoirs (or more in general with the environment) by means of quantum noise and studying the remaining degrees of freedoms within the formalism of the quantum open systems \cite{Petruccione}. Within the Markovian approximation this amounts to model reservoirs by means of Lindblad operators acting at the system ends and to replace the Liouville equation for the time evolution of the density matrix with a master equation of Lindblad type for the reduced density matrix acting of the Hilbert space associated to the degrees of freedoms of physical interest  \cite{Petruccione, PlenioJumps}.

The aim of the present paper is to use the formalism of the Lindblad master equation to provide  a first  characterization of the optimal transport in an open spin chain out of equilibrium  in terms of the von Neumann entropy (VNE). In this respect we consider a XXZ Heisenberg chain with twisted XY boundary reservoirs at the  ends which possesses  inversion symmetry with respect to the middle point of the chain. By means of this symmetry we derive a set of algebraic equations which allow to compute  in exact manner the stationary density matrix elements for chains of small sizes.  As a result we show that the maximal spin current always occurs  in proximity of a minimum  of the VNE and for particular choices  of parameters it can exactly coincide  with it. We also show that in  the strong coupling limit the minima of the VNE approach  zero (exactly vanishing for infinite coupling), this meaning  that the optimal transport is achieved in this case with states that are very close to pure states.

We remark that the open XXZ chain with an effective constant pumping at the chain ends which induce a gradient along the $z$ axis \cite{Casati2009} or with  easy-plane magnetizations at the boundaries \cite{PSS2012}, have also been investigated. To the best of our knowledge, however, no specific link between optimal transport and extrema of VNE has  been previously reported.

The paper is organized as follows. In Section II we present the model equation and discuss the quantum symmetry of the system. We show (in this section and in the  Appendix) how the symmetry operator can be used to derive a set of algebraic equations which allow to obtain elements of the stationary reduced density matrix in exact manner. In Section III this approach is used to investigate the relationship between spin current and optimal transport in terms of the  VNE for chains of small size. In the last Section the main results of the paper are briefly summarized.

\section{Model Equation and Symmetry Properties}
We consider the quantum Master equation for the open XXZ model in the Lindblad form:
\begin{equation}
\frac{\partial\rho}{\partial t}  = -i \left[  H,\rho\right]  + \gamma [-\frac{1}{2}
\sum_m\left\{  \rho,L_{m}^{\dag}L_{m}\right\}
+ \sum_m L_{m}\rho L_{m}^{\dag}] \label{lmeq}
\end{equation}
where  $\rho$ is the reduced density matrix, $H$ is the  $XXZ$ Hamiltonian
\begin{equation}
H=\sum_{k=1}^{N-1}\{\sigma_{k}^{x}\sigma_{k+1}^{x}+\sigma_{k}^{y}
\sigma_{k+1}^{y}+\Delta\sigma_{k}^{z}\sigma_{k+1}^{z} \}\label{Hamiltonian}%
\end{equation}
(we have fixed $\hbar=1$ and used $\dagger$  to denote adjoint operator)
and $L_{m}$ denote Lindblad operators introducing dissipation
in the system  at the ends of the chain. In the following we consider the case of
only two Lindblad operators of the form
\begin{equation}
L_1= \sigma_z(1)-i \sigma_x(1), \,\,L_2= \sigma_y(N) + i \sigma_z(N),
\label{lindbladL}
\end{equation}
with $\sigma_\alpha(i)$, $i=1,...N,\;  \alpha=x,y,z,$ denoting
usual Pauli matrices acting on site $i$, $N$ being the length of
the chain. This choice of Lindblad operators models  magnetic
reservoirs at the left ($i=1$) and right ($i=N$) end of the chain,
with fixed polarization along the axes $x$ and $y$, respectively.
Note that the operators  $L_{1}, L_{2} $, introduce  boundary gradient in
the polarization along the negative $x$ and positive $y-$direction
which allow to sustain a nonzero spin current in the transverse
z-direction which depends in a non trivial manner  on the
anisotropy parameter $\Delta$ (see also \cite{PSS2012},\cite{slava}).
Moreover, it is easy to check that with this choice of Lindblad operators the master
equation becomes invariant under the action of the symmetry operator
\begin{equation}
T= R \cdot P
\label{trot}
\end{equation}
where $R$ performs in the common $\sigma_j^z$ eigenbasis an inversion  of the spin  x-direction followed by a rotation in the $x-y$ plane, e.g.
\begin{equation}
R= \prod_{k=1}^N \substack{\otimes \\ _k}\left(
\begin{array}{cc}
 0 & i \\
 1 & 0 \\
\end{array}
\right),
\end{equation}
and the operator $P$ reverses the order of the sites $j \leftrightarrow N-j+1$, e.g.
$P (A_1 \otimes A_2 \otimes ...  \otimes A_N)= A_N \otimes A_{N-1} \otimes ...  \otimes A_1$, with $A_i$, $2\times 2$ matrices. Notice that $P$ is also performing  a reflection of the chain with respect to its middle point and as such, its action  is different for $N$ odd or $N$ even, since for $N$ odd the central site is unaffected by $P$ while for N even all lattice sites are affected (the reflection occurring with respect to an inter-site point).
Also, one can readily see that $T$ is a unitary operator $T^\dag=T^{-1}$ satisfying $T^2 = 1$ and
\begin{equation}
T L_2 = -i L_1 , \;\; T L_1 = i L_2.
\end{equation}
Moreover, $T$ commutes with  $H$ (it is a symmetry of the closed XXZ chain) as well as with the products  $L\dag _m L_m$, $m=1,2,$
\begin{equation}
[T, H]=0, \;\; [T, L^\dag_m L_m]=[T, L_m L^\dag_m]=0, \; m=1,2,
\end{equation}
this making $T$, for the chosen Lindblad operators (\ref{lindbladL}), a symmetry of the open XXZ chain.

In terms of stationary density matrix $\rho_s$, the above  relations, together with  Eq. (\ref{lmeq}) and the assumption of uniqueness
 for $\rho$ (verifiable as in \cite{ProsenUniqueness}), imply that $[\rho_s,  T]=0$, or equivalently:
\begin{equation}
T \rho_s T^{-1} = \rho_s.
\label{trota}
\end{equation}
This property can be used to obtain exact unique solutions of the stationary Lindblad master equation for arbitrary values of the parameters $\Delta, \gamma$. In this respect, it is worth to note that the stationarity of (\ref{lmeq}) together with  the fulfillment  of Eq. (\ref{trota}) and the condition $Tr(\rho)=1$,  provide a system of $d_N$ algebraic equations for matrix elements $\rho_{ij}$, with $d_N$ given by
\begin{equation}
d_N=\left\{
\begin{array}
[c]{c}%
2^{2 N -1}   \text{\;\;\;\;\;\;\;\;\;\;\;\; for}\; \text{\ $N$} \text{\ odd}, \\
(1+2^N) 2^{N -1}  \text{\;\; for}\; \text{\ $N$} \text{\ even}.
\end{array}
\right. \label{dim}
\end{equation}
Thus, for example, for  the case $N=3$, one can easily derive from Eq. (\ref{trota}) that  the most general form  for the matrix $\rho$, compatible with the symmetry $T$, is the matrix with real diagonal elements given by
\begin{eqnarray}
& &\rho_{11}=\rho_{88}= a_{11}, \;\rho_{22}=\rho_{44}=a_{22}, \nonumber \\
& &\rho_{33}=\rho_{66}=a_{33}, \;\rho_{55}=\rho_{77}= a_{55},
\label{elem1}
\end{eqnarray}
and with off-diagonal elements $\rho_{i,j}=\rho_{j,i}^*=a_{i,j} +i b_{i,j}$  satisfying the relations:
\begin{eqnarray}
& & \rho_{28}=-\rho_{14}^*, \;\rho_{34}=i\rho_{26}^*, \;\rho_{38}=-\rho_{16}^*, \nonumber \\
& & \rho_{45}=-i\rho_{27}, \;\rho_{46}=i\rho_{23}^*, \;\rho_{47}=\rho_{25}, \nonumber \\
& & \rho_{48}=i\rho_{12}^*, \;\rho_{56}=i\rho_{37}^*, \;\rho_{58}=-\rho_{17}^*, \nonumber \\
& & \rho_{67}=\rho_{35}, \;\rho_{68}=i\rho_{13}^*, \;\rho_{78}=i\rho_{15}^*, \nonumber \\
& &\rho_{18}=(1+i) a_{18}, \;\rho_{24}= (1+i) a_{24}, \nonumber \\
& &\rho_{36}= (1+i) a_{36}, \;\rho_{57}=(1+i) a_{57},
\label{elem2}
\end{eqnarray}
where the star denotes the complex conjugation and $a_{ij}, b_{ij}$, are real numbers (see details in the Appendix).
By substituting this form of $\rho$ into Eq. (\ref{lmeq}) and requiring stationarity, one gets a system  of $32$ independent equations for $a_{ij}, b_{ij}$ which admits a unique solution when the normalization condition $Tr[\rho]= 2(a_{11}+ a_{22}+a_{33}+ a_{55})=1$  is imposed. We remark that while for interacting closed many body  systems a complete characterization of the reduced density matrix in terms of the system symmetries is possible (mainly  for  systems with the permutational invariance \cite{RDM-perm}), very little is known in this respect for  open quantum systems.

\begin{figure}[ptb]
\centerline{
\includegraphics[height=4.5cm]{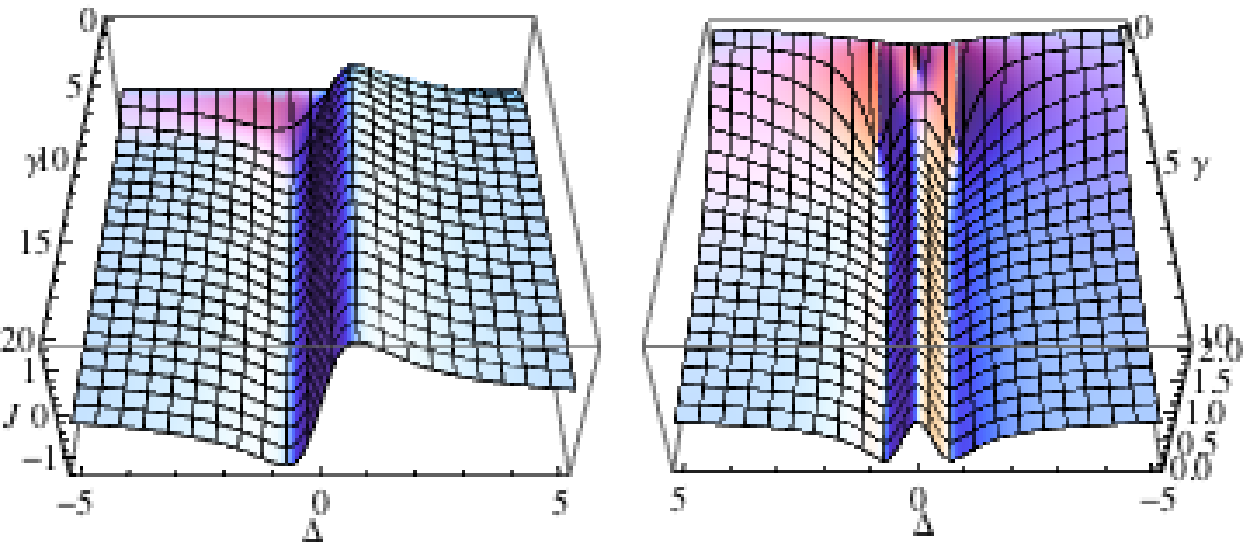}
}
\vskip -1.cm
\centerline{
\includegraphics[width=5. cm,height=7.cm,clip]{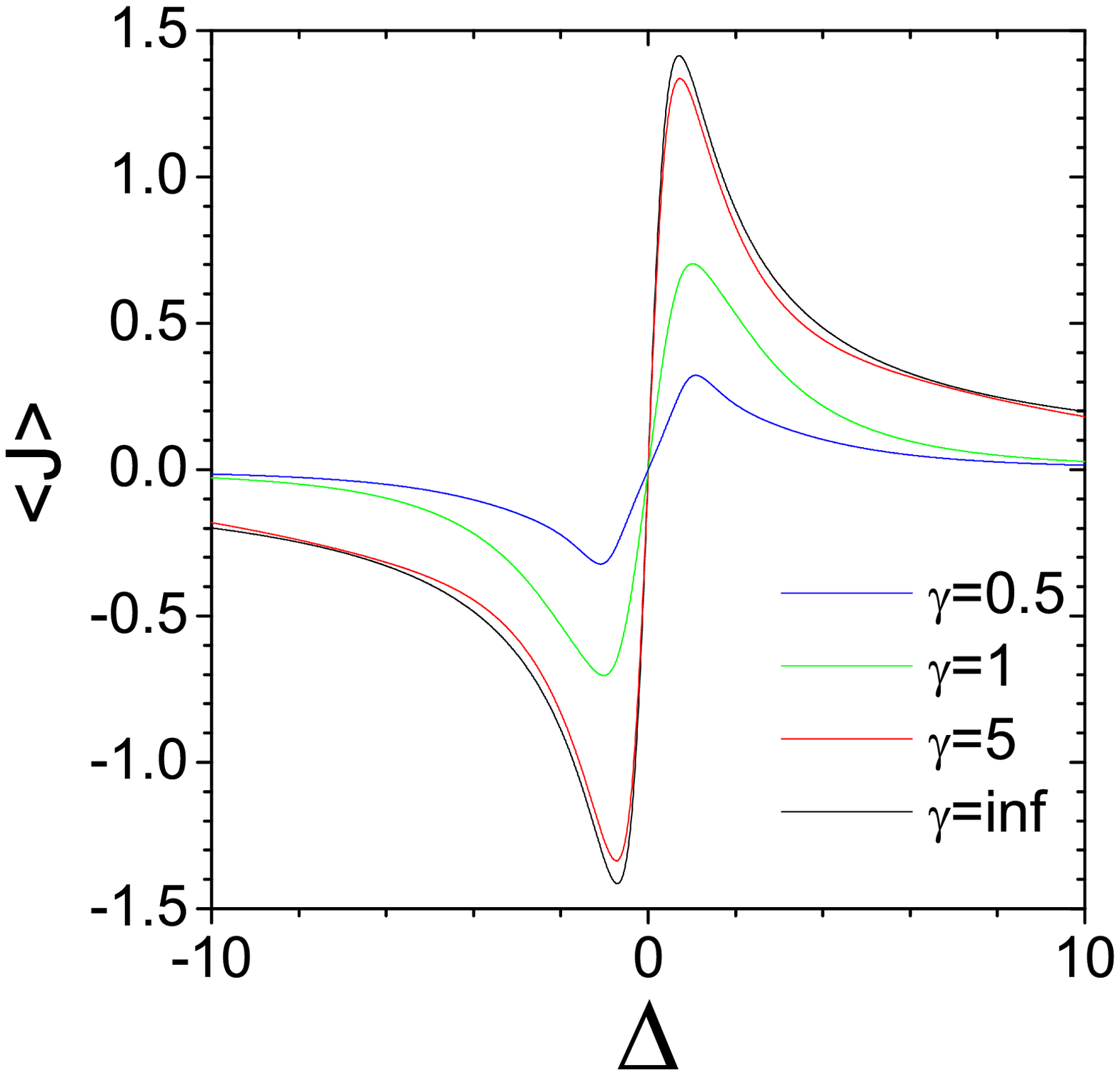}
\hskip -1.cm
\includegraphics[width=5. cm,height=7.cm,clip]{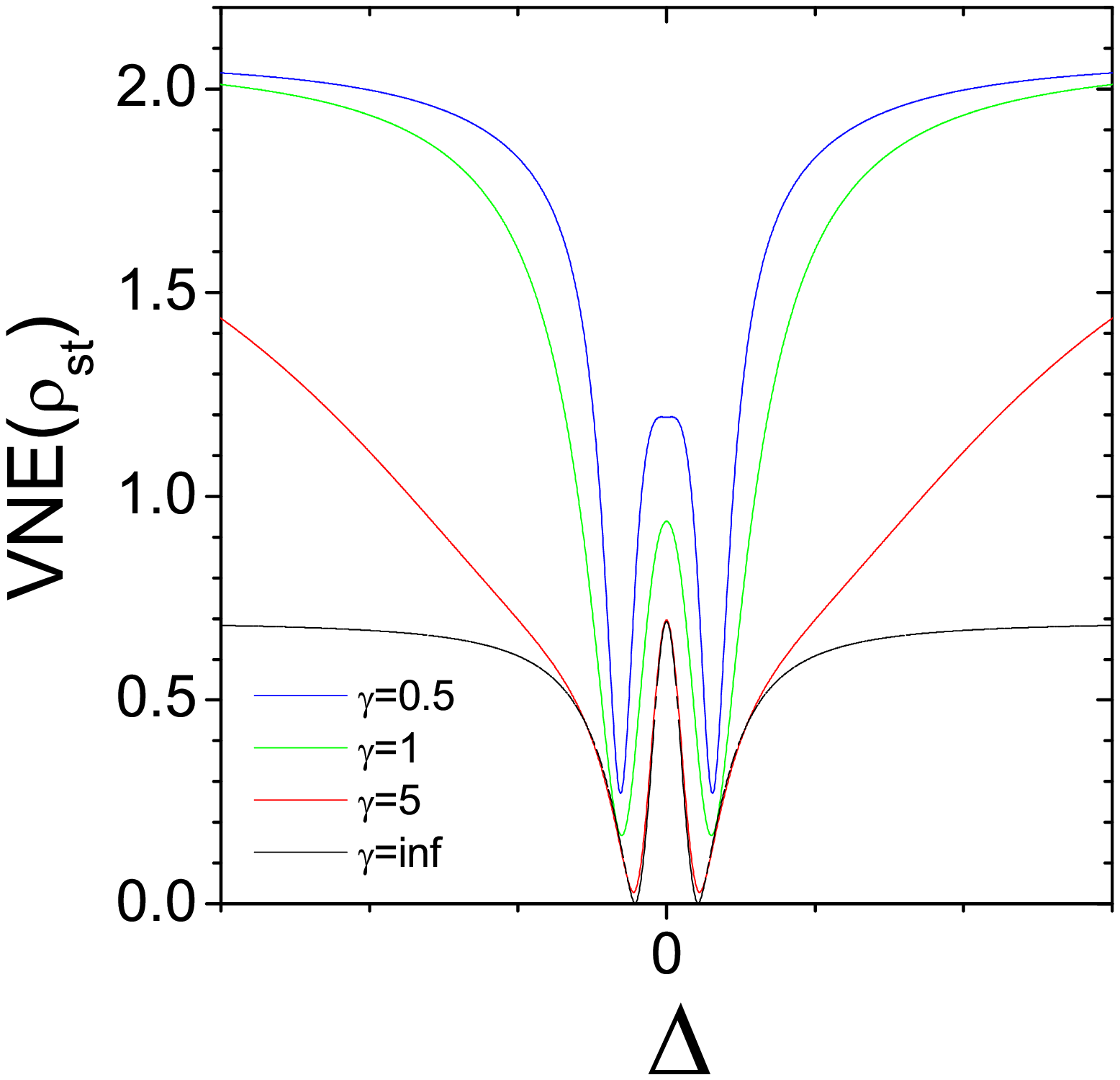}
}
\vskip -2.5cm
\caption{(Color online) Top panels. Stationary spin current  $J$ (left) and von Neumann entropy (right) as a function of parameters  $\gamma, \Delta$ for the open
$\mathcal{H}_{XXZ}$ chain  with Lindblad operators (\ref{lindbladL}) and length $N=3$. Bottom panels. Sections of the spin current  $J$ (left) and von Neumann entropy (right) surfaces in the top panel taken at different values of  parameter  $\gamma$.}
\label{Fig1}%
\end{figure}
\begin{figure}[ptb]
\centerline{
\includegraphics[width=8.6 cm,height=10.6cm,clip]{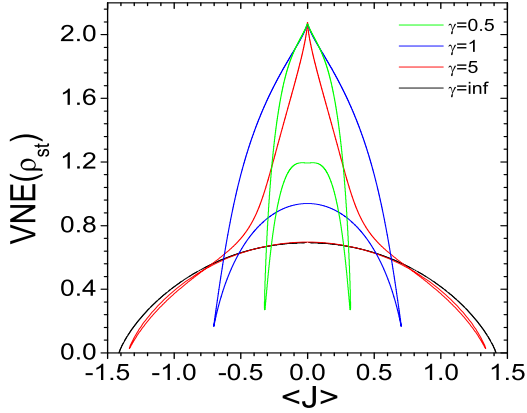}
}
\vskip -3.5cm
\caption{(Color online) Average spin current vs von Neumann entropy for different values of the coupling parameter $\gamma$. The curves are obtained as parametric plots for fixed $\gamma$ and $\Delta$ varied in the range  $]-\infty, \infty[$. The VNE attains it maximum value (same for all curves) at  $J=0$ and for $\Delta=\pm \infty$.}
\label{Fig2}%
\end{figure}

Although our approach is completely general, the rapid exponential
 growth of the number of algebraic equations to solve as  $N$ is increased see
  Eq. (\ref{dim}), quickly restricts the application of the symmetry  method to
  chains of small sizes. On the other hand,  qubits systems with few sites can be
  already of potential interest for applications, and in view  to the scarcity of
   exact results for open systems, we shall  take the opportunity to investigate
   transport properties of these systems in exact manner for arbitrary parameters.
   In this respect we remark that this problem has been also investigated in previous
   papers  using quantum trajectory method \cite{Casati2009} or direct  time integrations of
   the Lindblad master equation \cite{PSS2012}. Our approach, being exact,  is free of
   convergence problems of time integrations present in these alternate approaches
   (typically the convergence becomes very slow in the critical region  $-1\le \Delta \le 1$ and
    in the weak and strong  coupling limits).

In addition, the steady state density matrices for positive and
negative values of the anysotropy parameter $\Delta$ are connected
via relations \cite{slava}
\begin{equation}
\rho_{s}(N,-\Delta)=U\rho_{s}^{\ast}(N,\Delta)U\text{ for }N\text{
even.}
\label{SymmetryEven}
\end{equation}
\begin{equation}
\rho_{s}(N,-\Delta)=(\sigma^{y})^{\otimes_{N}}U\rho_{s}^{\ast}(N,\Delta
)U(\sigma^{y})^{\otimes_{N}}\text{ for }N\text{
odd}\label{SymmetryOdd}
\end{equation}
where $U=\sigma^{z} \otimes I \otimes \sigma^{z}  \otimes ....$ is a
tensor product over all odd sites, which has the property $U
H(\Delta) U= -H(-\Delta)$, and $\rho^{\ast}$ a complex conjugated
matrix in the basis where $\sigma^{z}$ is diagonal. Note that the
above transformations depend on parity of $N$ and, in particular,
result in the current being an odd (even) function of $\Delta$ for
odd (even) $N$,
\begin{equation}
j(\Delta)=(-1)^{N} j(-\Delta) \label{SymmetryCurrent}
\end{equation}
For the case $N=3$ the system of $d_3=32$ algebraic equations obtained with the help of the symmetry
(\ref{trota})  is reported in Appendix A.
From the  solution of this system one can easily check that the analytical expression for the average current $J(\gamma,\Delta) = Tr(\rho (\sigma_x(i) \sigma_y(i+1)-\sigma_y(i)\sigma_x(i+1)))$ is of the form
\begin{equation}
J(\gamma, \Delta)= \frac{\sum_{i=0}^{N^2+1} \alpha_i(\Delta) \gamma^{2 (i+1)}}{\sum_{i=0}^{N^2+2} \beta_i(\Delta) \gamma^{2 i}}, \;\;\; N=3,
\label{current}
\end{equation}
with coefficients $\alpha_i(\Delta), \beta_i(\Delta)$, odd and
even polynomials in  $\Delta$, respectively. This implies that the
current is an odd function of $\Delta$, in accordance to
(\ref{SymmetryCurrent}). The first leading coefficients relevant
for strong and weak coupling limits are
\begin{eqnarray}
&&\alpha_{0}(\Delta)= 144 \Delta^5 (\Delta^2-1)^4 (9 \Delta^2 + 59),\;\nonumber \\
&&\alpha_{1}(\Delta)= 8 \Delta^3 (\Delta^2-1)^2 (9420 + 13895 \Delta^2+ 5822 \Delta^4 \nonumber \\
&& + 6561 \Delta^6+ 2396 \Delta^8 + 186 \Delta^{10} ) , \nonumber \\
&&\;\;\;\;\;\;\;   ... \nonumber \\
&&\alpha_{9}(\Delta)= 248832 \Delta (265 + 96 \Delta^2) , \nonumber \\
&&\alpha_{10}(\Delta)= 8957952 \Delta.\;\nonumber \\
\end{eqnarray}
and
\begin{eqnarray}
&&\beta_{0}(\Delta)= 24 \Delta^4 (\Delta^2-1)^4 (416+155\Delta^2+9\Delta^4),  \nonumber \\
&&\beta_{1}(\Delta)= 2 \Delta^2 (\Delta^2-1)^2 (34032 + 45152 \Delta^2 - 7764 \Delta^4 \nonumber \\
&&\;\;\;\;\;\;\;  ... \nonumber \\
&&\beta_{10}(\Delta)= 62208 (228 \Delta^4 + 644 \Delta^2 + 355), \nonumber \\
&&\beta_{11}(\Delta)= 2239488 (1+2 \Delta^2). \nonumber \\
\end{eqnarray}
\begin{figure}[ptb]
\centerline{
\includegraphics[width=5.6 cm,height=7.6cm,clip]{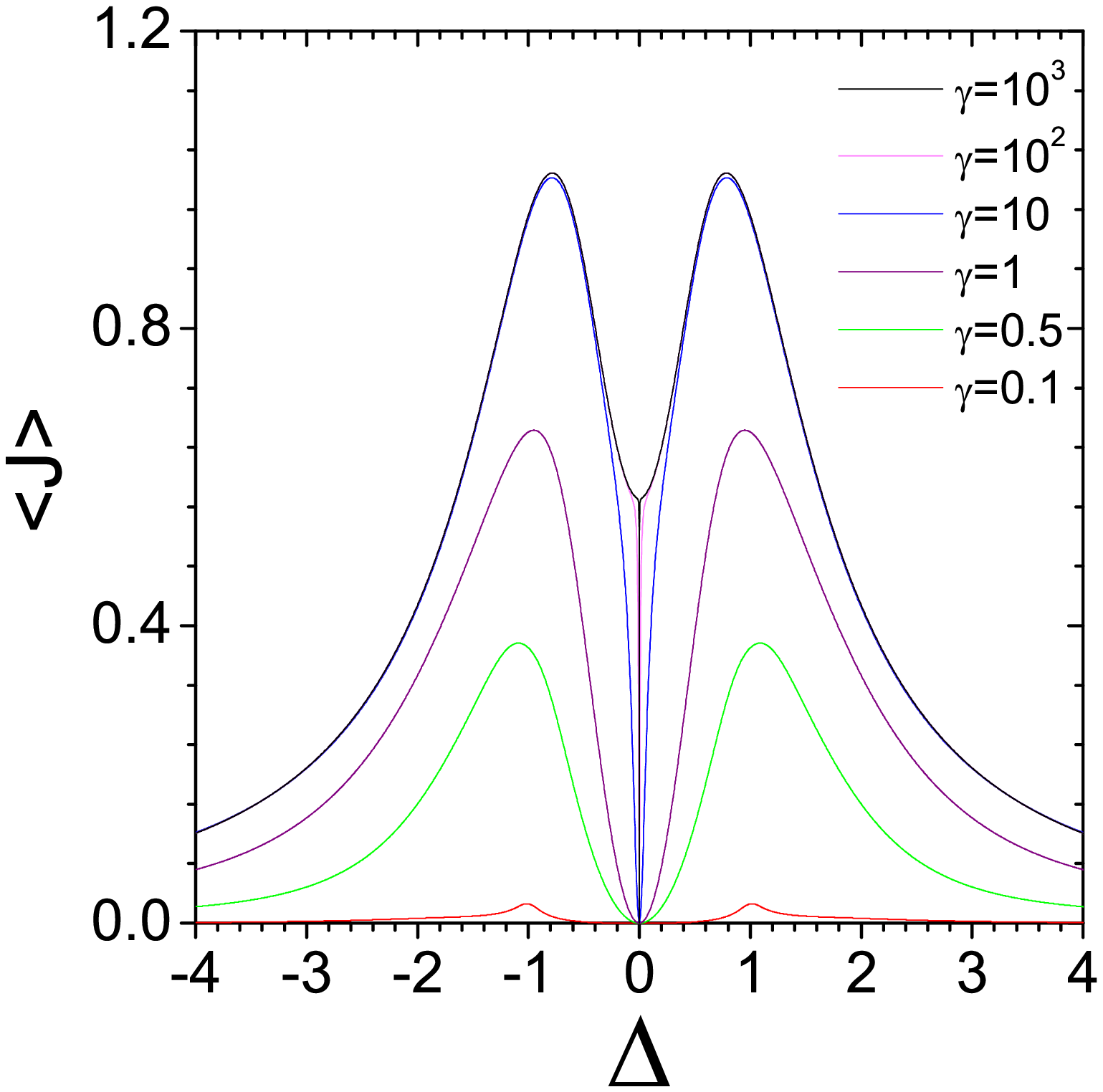}
\hskip -1.2cm
\includegraphics[width=5.6 cm,height=7.6cm,clip]{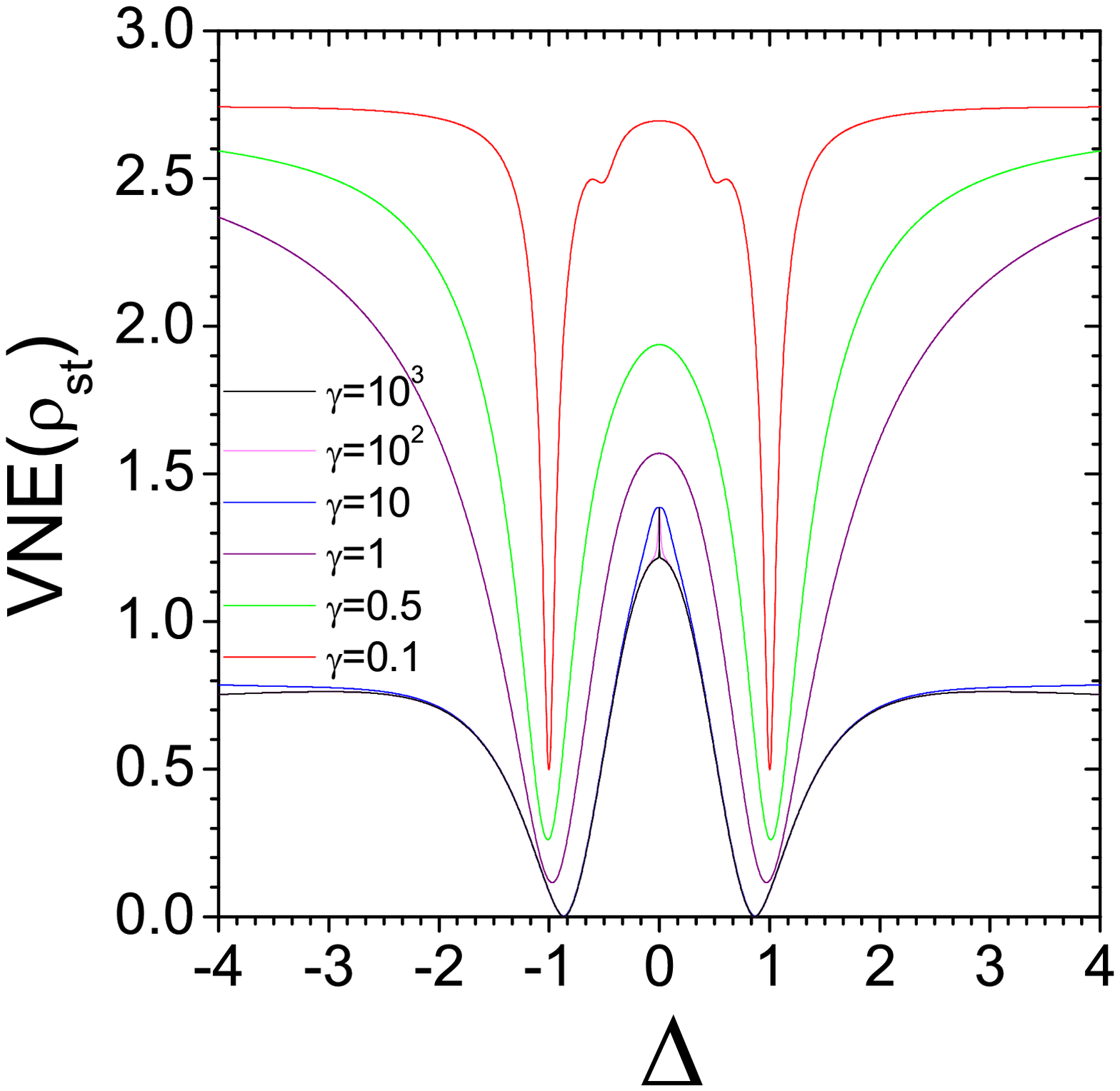}
}
\vskip -3cm
\centerline{
\includegraphics[width=5.6 cm,height=7.6cm,clip]{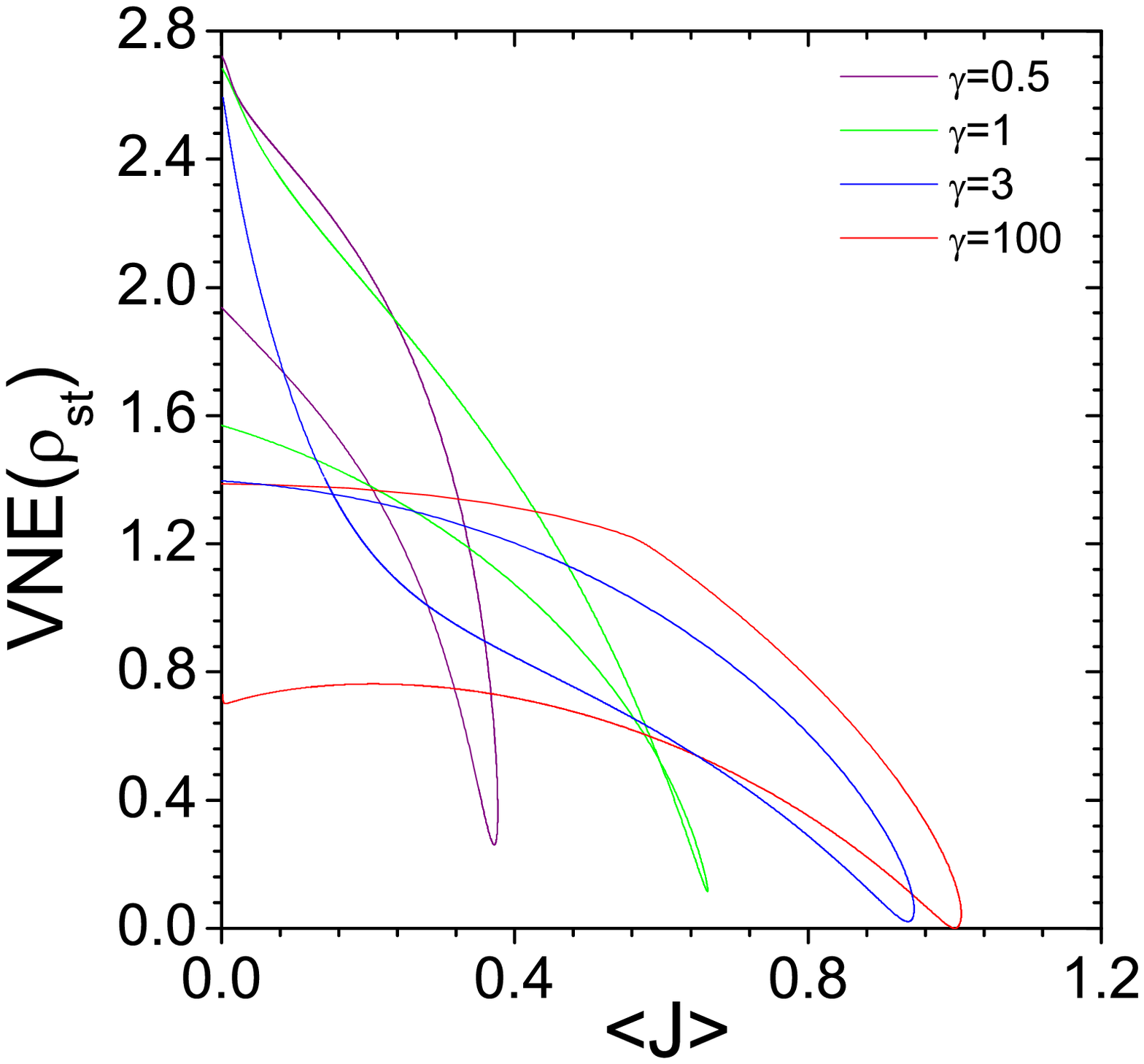}
\hskip -1.2cm
\includegraphics[width=5.6 cm,height=7.6cm,clip]{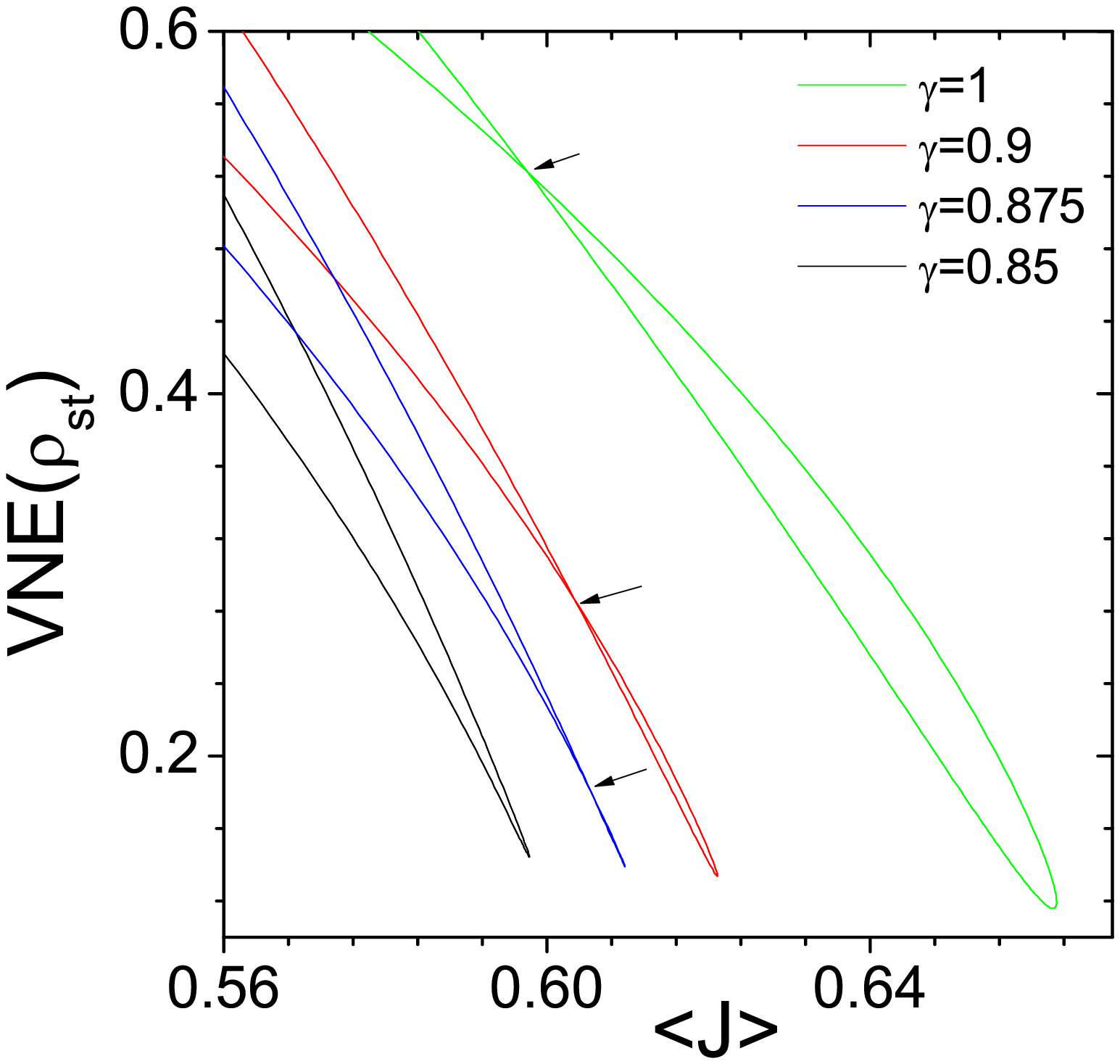}
} 
\vskip -2.8cm \caption{(Color online) Top panels.Stationary spin
currents  $J$ (left panel) and von Neumann entropies (right panel)
vs $\Delta$ for the $\mathcal{H}_{XXZ}$ chain of length $N=4$ for
different values of $\gamma$ indicated in the figure. Curves are
obtained from exact stationary solution of the Lindblad Master
Equation. Bottom panels. Corresponding von Neumann entropies vs
stationary currents $J$ for different coupling constants $\gamma$
indicated in the panels. Bottom right panel show details of the
crossover through  the cusp point at which  minimum of VNE and
maximum of current are in  exact coincidence (e.g. they occur  at
the same value of $\Delta$). All curves are obtained from exact
stationary solutions of the Lindblad equation.}
\label{Fig3}%
\end{figure}

To avoid  lengthy expressions we omitted other  coefficients since they can be easily obtained from the exact solution of the algebraic system in  Appendix A. Substituting the leading coefficients  into Eq. (\ref{current}) we obtain expressions for the current
\begin{eqnarray}
&&
J(\Delta,\gamma)\approx\frac{4\gamma^2\Delta[18(9\Delta^2+59)(\Delta-\Delta^3)^2 +\gamma^2 F_1]}{12 \Delta^2(\Delta^2-1)^2(9\Delta^4+155\Delta^2+416)+\gamma^2 F_2}, \nonumber \\
&& J(\Delta,\gamma)\approx\frac{4\Delta(96\Delta^2+36\gamma^2+265)}{228\Delta^4+36\gamma^2(1+2\Delta^2)+664 \Delta^2+355},
\label{Jleading}
\end{eqnarray}
valid for arbitrary values of $\Delta$ and small, $\gamma \ll 1$, and large, $\gamma \gg 1$,  values of $\gamma$, respectively. Here $F_1=186\Delta^{10}+2396\Delta^8+6561\Delta^6+5822\Delta^4+13895\Delta^2+9420$ and
$F_2=153\Delta^{12}+3411\Delta^{10}+19179\Delta^8+24007\Delta^6+7764\Delta^4+45152\Delta^2+34032$.
Note, from the second Eq. in (\ref{Jleading}),  that in the infinite coupling limit $\gamma \rightarrow \infty$ the current is:
\begin{equation}
\lim_{\gamma\rightarrow \infty} J(\gamma,\Delta)= \frac{4 \Delta}{1 + 2 \Delta^2},
\end{equation}
this coinciding with the expression  derived in \cite{slava} by
means of a perturbative approach. Similar results are obtained for
other values of $N$, with the only difference that the case of $N$
even the coefficients of the series in Eq. (\ref{current}) are
even functions of $\Delta$, e.g. for even (odd) $N$ the current is
an even (odd) function of $\Delta$. Explicit expressions of $\rho$
(and derived quantities), however, become very involved as $N$
increases since the number of algebraic equations to solve
increases exponentially (see Eq. (\ref{dim}).
 Although  it may be a problem to get analytical expressions for such large systems, it is not a problem to solve them
  numerically with  very high accuracy to be considered exact for any practical
  purpose.

\section{Transport properties and von Neumann entropy for out of equilibrium finite size XXZ chains}
The symmetry approach discussed above allows to obtain exact
analytical  results for the physical quantities of interest in the
transport,  such as the spin current $J$ and the von Neumann
entropy $VNE(\rho_{st})=-\sum_i \lambda_i \ln \lambda_i$,  where
$\lambda_i$ denote the eigenvalues of the stationary reduced
density matrix.
\vskip 1cm
\begin{figure}[ptb]
\centerline{
\includegraphics[width=5.6 cm,height=7.6 cm,clip]{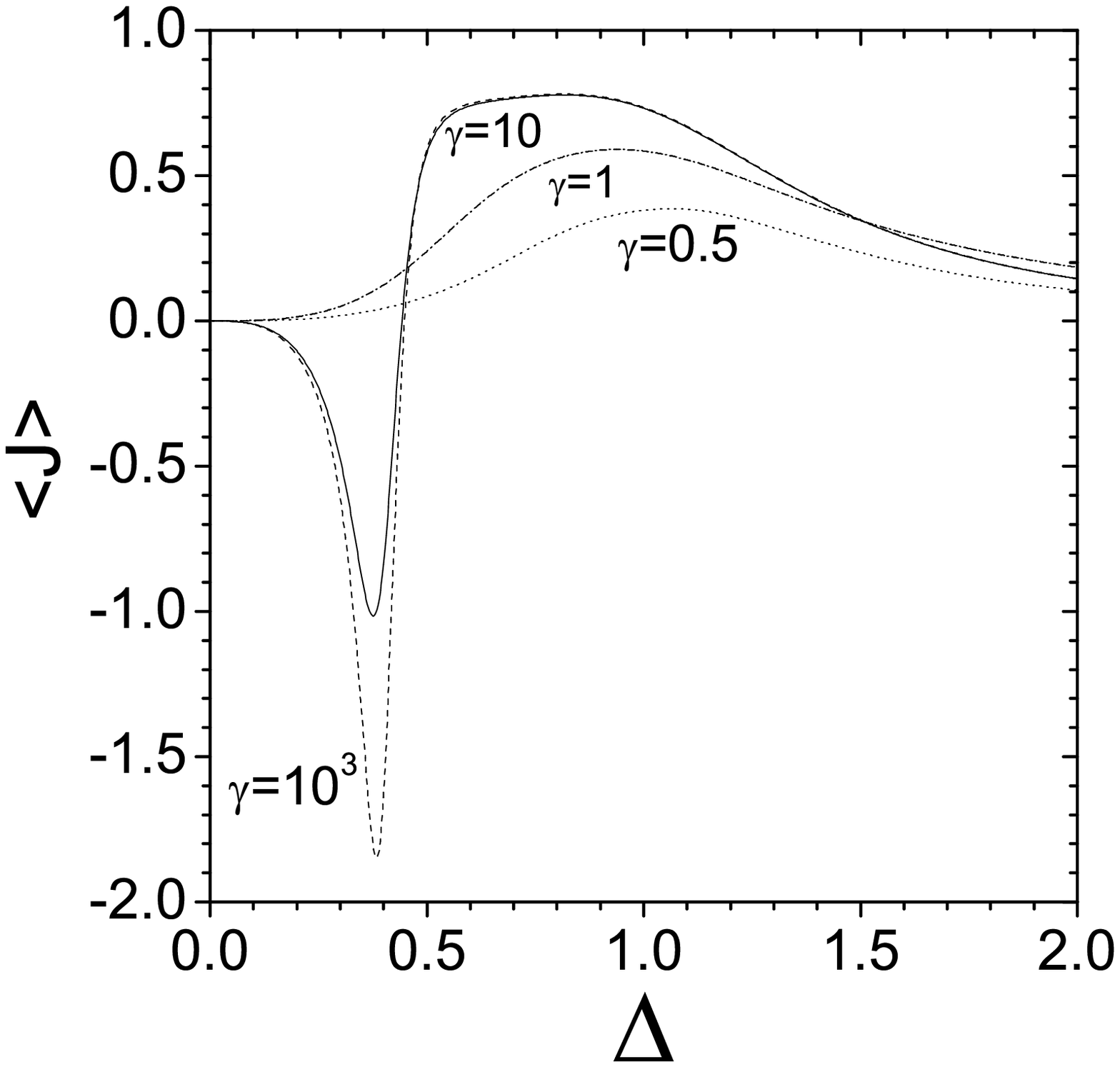}
\hskip -1.2cm
\includegraphics[width=5.6 cm,height=7.6 cm,clip]{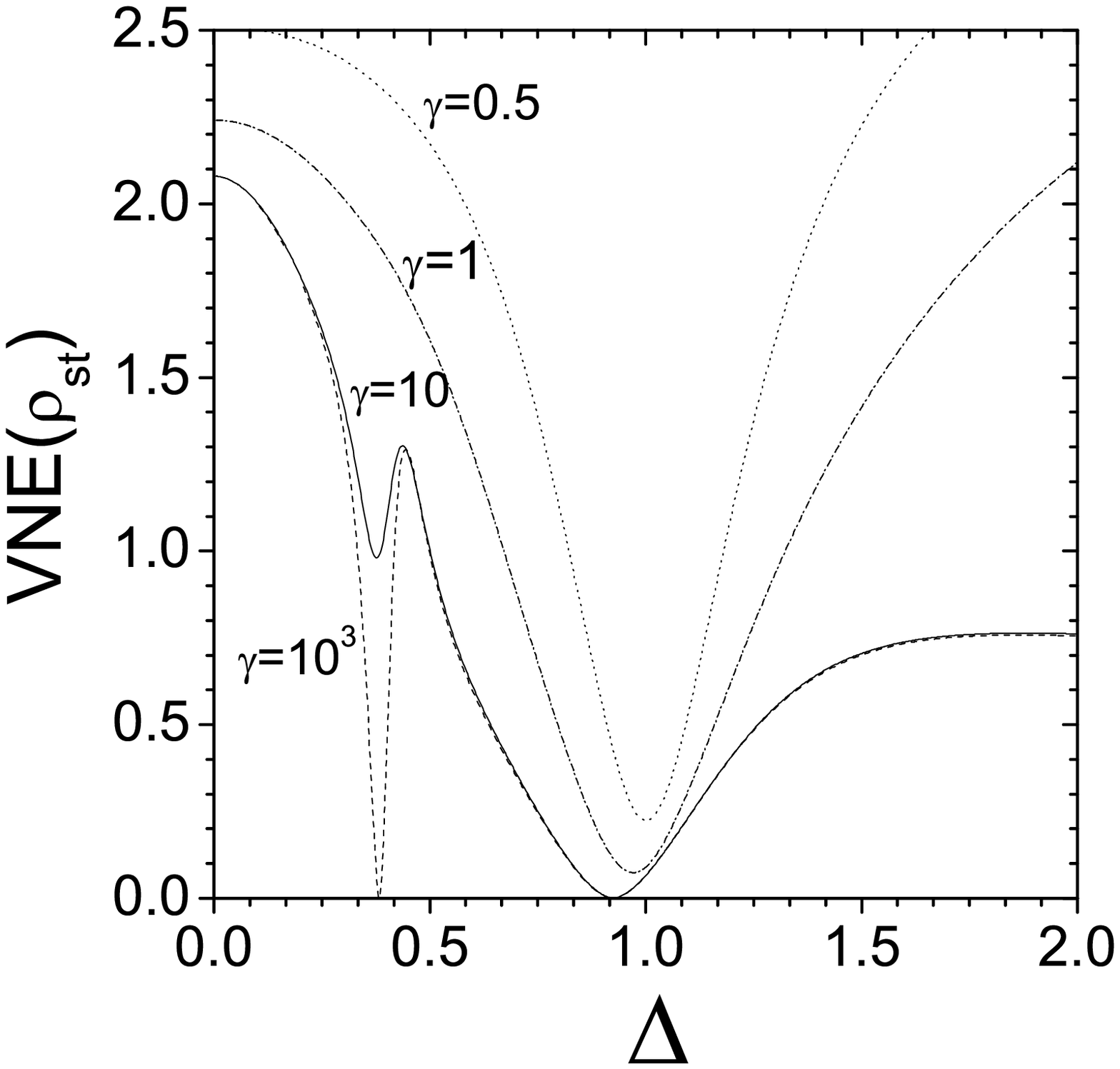}
}
\vskip -2.3cm
\centerline{
\includegraphics[width=5.6 cm,height=7.6 cm,clip]{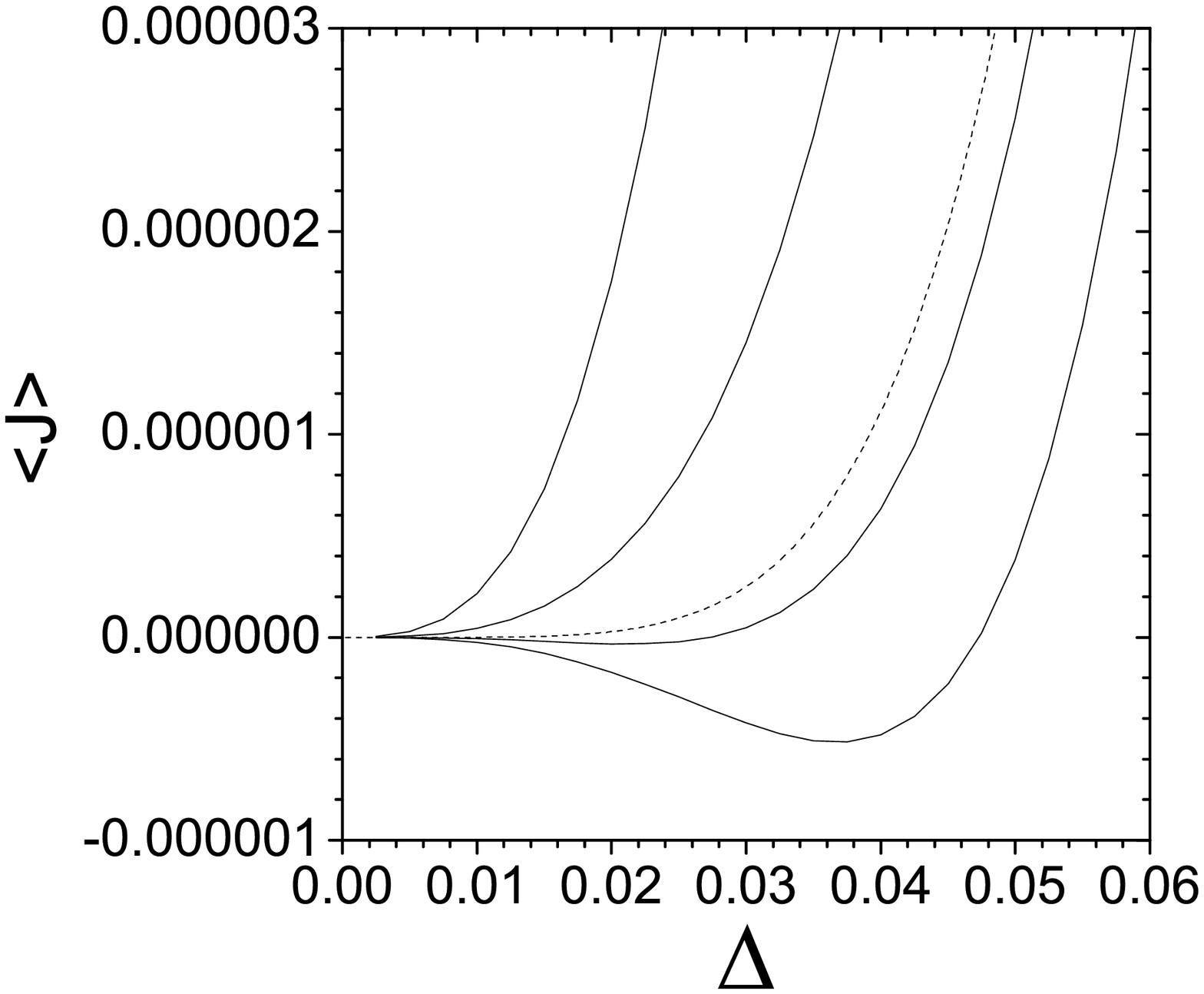}
\hskip -.6cm
\includegraphics[width=5.6 cm,height=7.6 cm,clip]{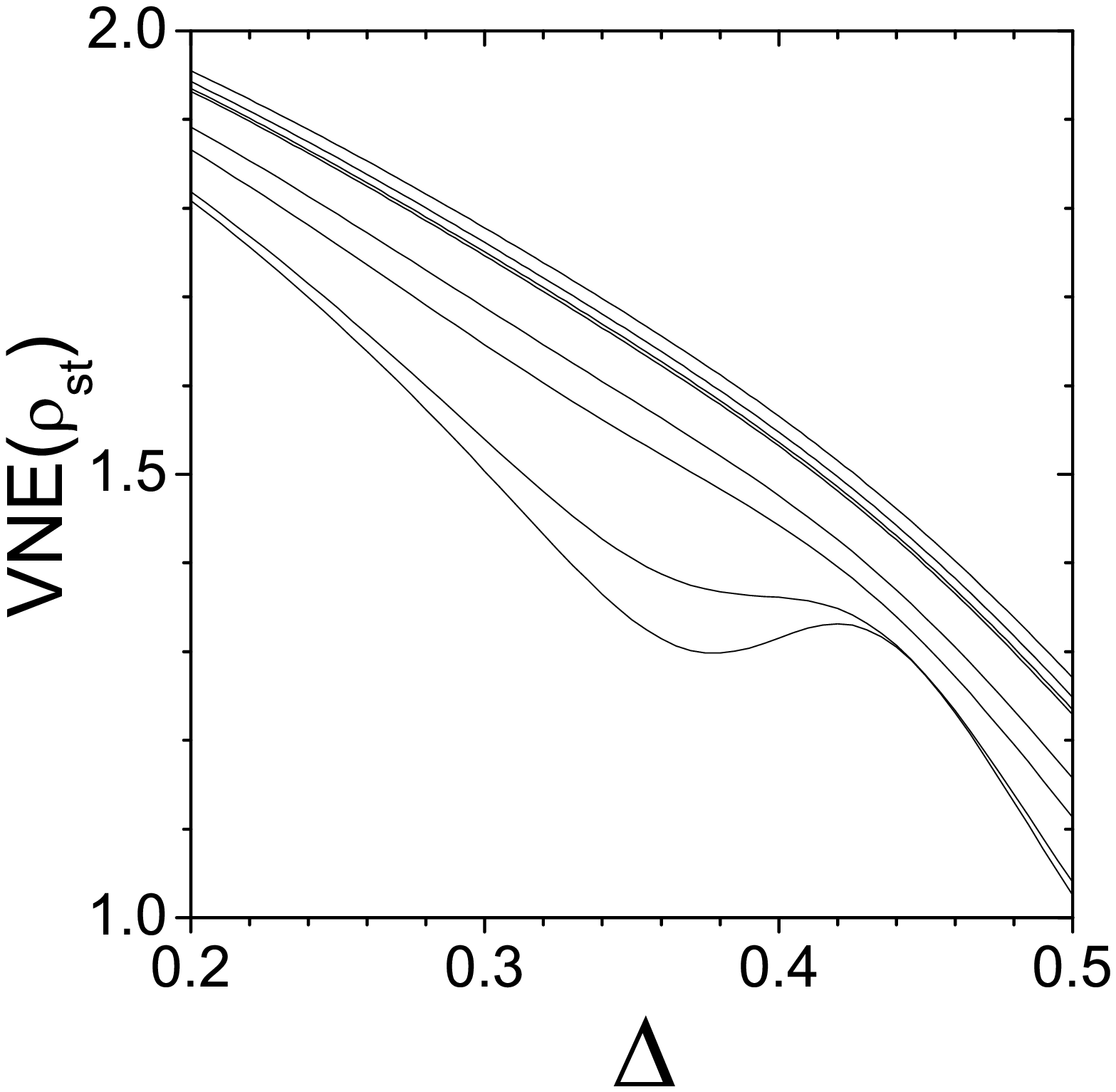}
}
\vskip -2cm
\caption{(Color online) Top panels. Stationary spin currents  $J$ (left panel) and von Neumann entropies (right panel)   vs $\Delta$ for the
$\mathcal{H}_{XXZ}$ chain of length $N=5$ for different values of $\gamma$ as indicated in the figure.
Curves are obtained from exact stationary solutions of the Lindblad Master Equation and, taking into account their antisymmetry property in $\Delta$,  they have been plotted  for positive $\Delta$ values only, just for graphical convenience). Bottom Panels. Details of the reversal current transition (left  panel) and corresponding von Neumann entropy. Curves from left to right  in the left panel refer to $\gamma$ values $1.9, 1.95, 1.96285, 1.965, 1.97$, respectively. Curves from top to bottom  in the right panel refer to $\gamma$ values $1.8, 1.9, 1.97, 2, 2.5, 3, 5 , 6$, respectively.
 }
\label{Fig4}%
\end{figure}

In the top panels of Fig. \ref{Fig1} we show stationary spin current  $J$ (left) and von Neumann entropy (right), as obtained from the exact solution of the system in Eq. (\ref{systemn3}) for the case $N=3$, as function functions of the parameters  $\gamma, \Delta$. Sections of the spin current  $J$ (left) and von Neumann entropy (right) surfaces are reported in the bottom panels for different values of  parameter  $\gamma$. We see that the mean current is an odd function of $\Delta$ and the VNE curve has minima very close (practically in correspondence) to maxima of $|J|$. This is clear also from Fig. \ref{Fig2} where a parametric curve J vs VNE is reported for different values of $\gamma$. Notice that in the strong coupling limit the VNE becomes zero exactly in correspondence to the maxima of  $|J|$. This implies that optimal transport (e.g. the maximal current)  is achieved in correspondence of pure states of the open $XXZ$ chain . Notice that the current vanishes for $\Delta=\pm \infty$ (Ising limit) and the VNE attains its maximum at $\Delta=0$ (free fermion point).
It is also worth to note here that although analytic results are available for the
Lindblad Master equation at the free fermion point \cite{Pros08,ZnidaricJStat2010, ZnidaricJPhysA, KarevskiPatini, ProsenZnidaric}, these results do not apply to our choice of Lindblad operators (\ref{lindbladL}).

Similar results, obtained for the case $N=4$, are shown in Fig. \ref{Fig3}. We see that the spin currents
 $J$ (top left panel) for different values of $\gamma$ are even functions of $\Delta$, and that the von Neumann
 entropies (top right panel) attain their maximal value at $\Delta=\pm \infty$ where $J$ is zero, while their
  minima are still very close to the maxima of $|J|$. Notice from the bottom left panel, that while the minima
  of the VNE always reduce to zero in the strong coupling limit $\gamma \rightarrow \infty$, the correspondence
  between the max of $|J|$ and the min of VNE remains  non exact also in this limit.
This property is completely general for all $N > 3$ (see below).
It is also worth  to note from this Fig. that as $\gamma$ is
increased from small (green curve) to strong coupling limit (red
curve) the  VNE-$J$ curve
 undergoes a crossing point with the development of a cusp at a particular value of $\Delta$ for which the
  maximal current exactly coincides with the minimum of the VNE. This process is clearly shown in the bottom
  right panel of Fig. \ref{Fig3} where the cusp appears  on the black curve $\gamma=0.85$ for $J\approx 0.6$.
  This folding of the curve as $\gamma$ is increased is also a general property observed for other values of $N$.
\begin{figure}[ptb]
\centerline{
\includegraphics[width=5.6 cm,height=7.6 cm,clip]{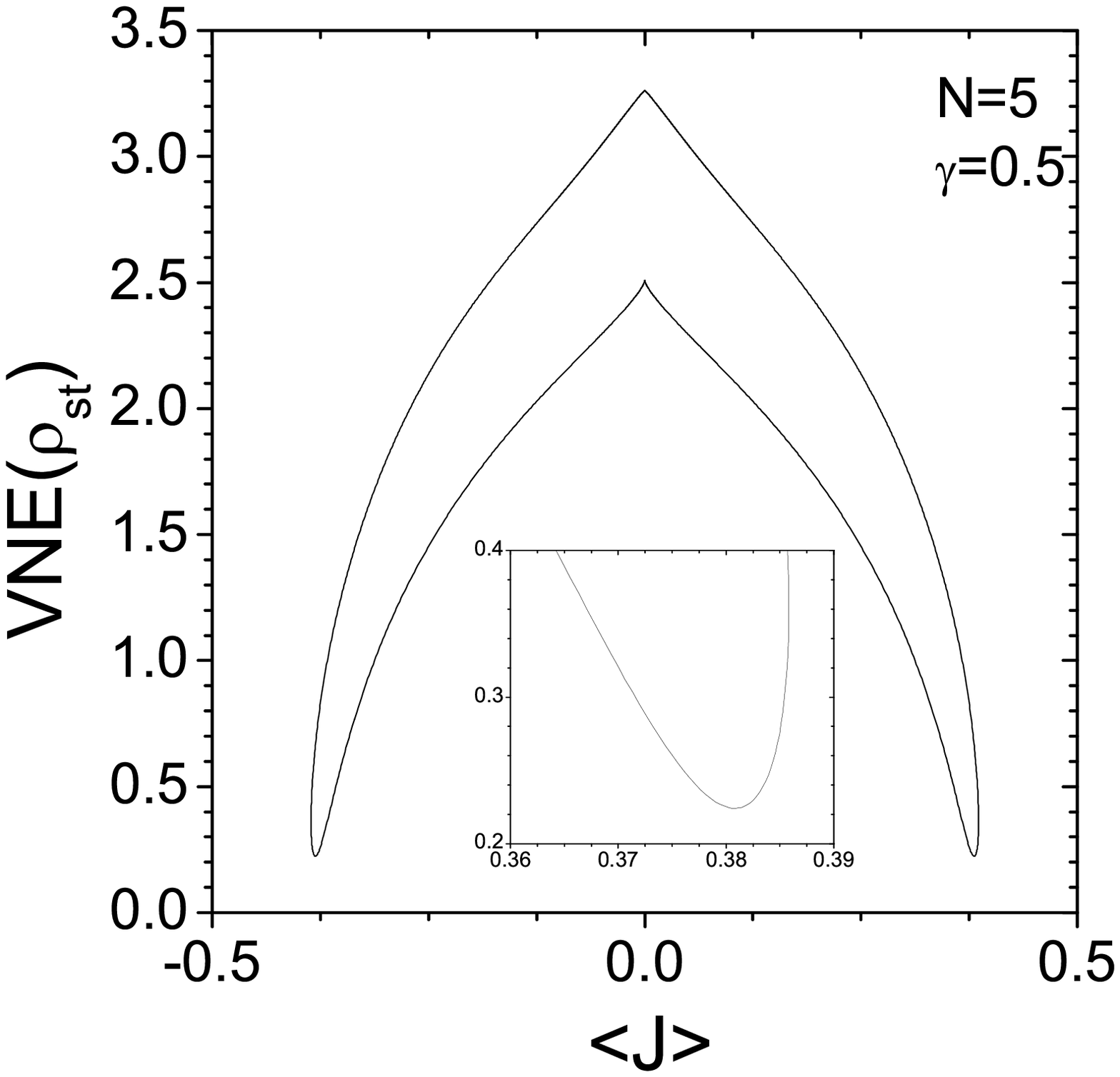}
\hskip -1.2cm
\includegraphics[width=5.6 cm,height=7.6 cm,clip]{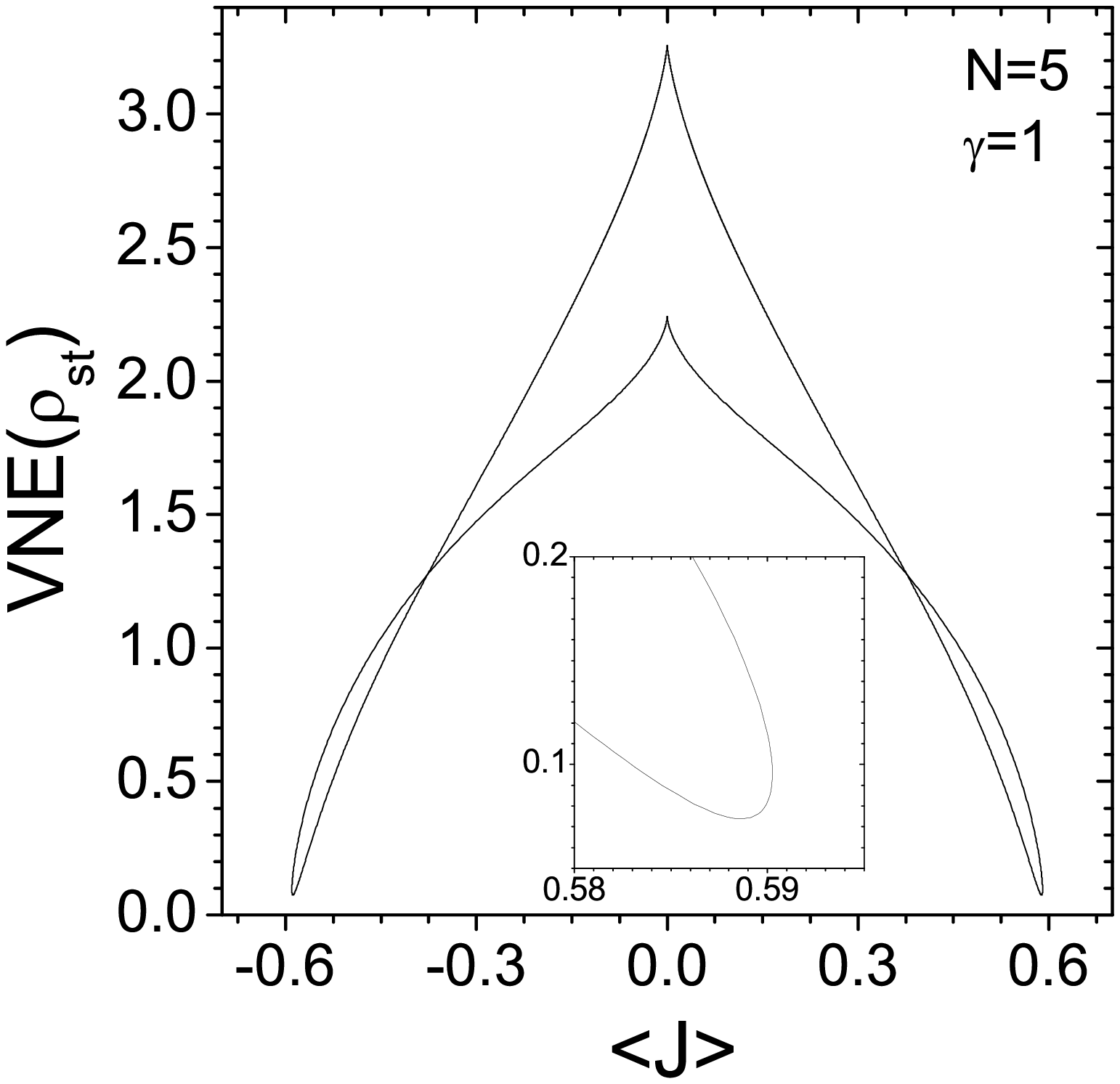}
}
\vskip -2.3cm
\centerline{
\includegraphics[width=5.6 cm,height=7.6 cm,clip]{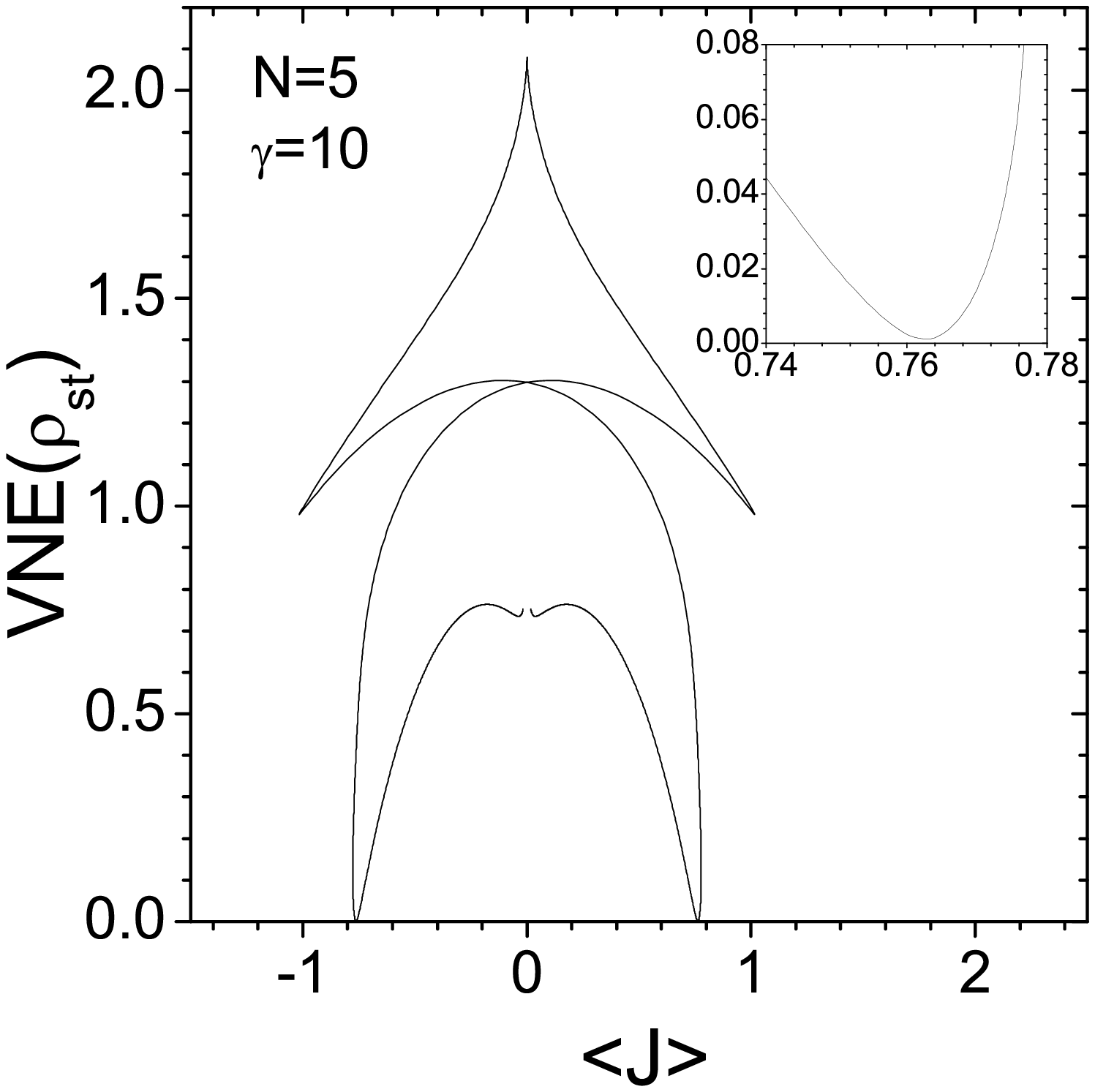}
\hskip -1.2cm
\includegraphics[width=5.6 cm,height=7.6 cm,clip]{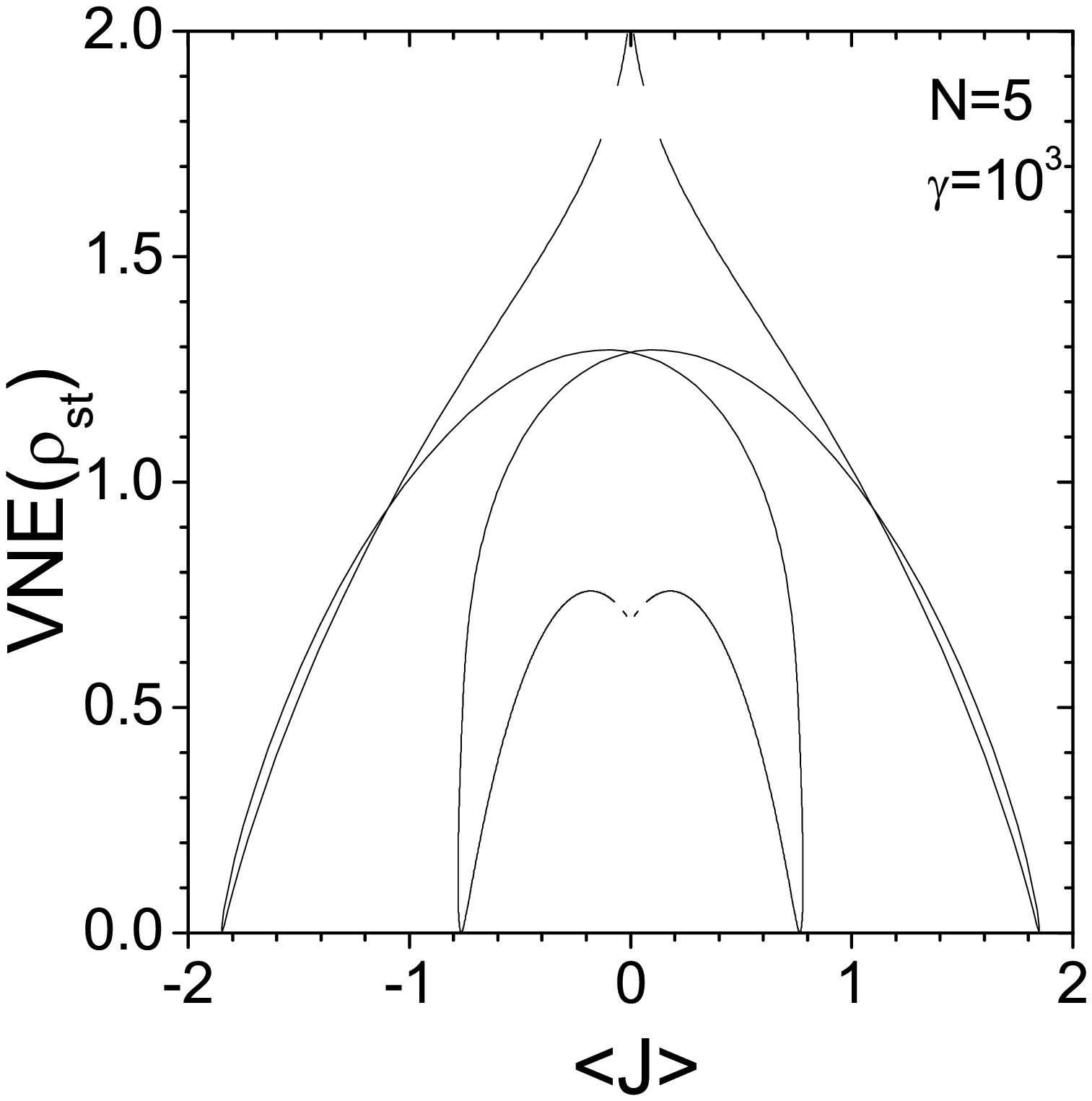}
} 
\vskip -2cm \caption{(Color online) The von Neumann entropy VNE
vs stationary current $J$ for the open $\mathcal{H}_{XXZ}$ chain
of length $N=5$ with different coupling constants $\gamma$ as
indicated in the panels.The curves are obtained as parametric
plots for fixed $\gamma$ and $\Delta$ varied in the range
$]-\infty, \infty[$. From panel insets one can see that the minima
of the VNE are always very close to  maxima of the absolute value
of the current. As the coupling constant $\gamma$ is increased the
values of the VNE at the minima decrease and in  the strong
coupling limit $\gamma \rightarrow \infty$ they reduce  exactly to
zero (pure states). }
\label{Fig5}%
\end{figure}

As $N$ is increased, however, curves becomes more and more
complicated with the occurrence of the interesting phenomenon of
current reversal. This is first observed for the case $N=5$
reported in Figs. \ref{Fig4}, \ref{Fig5}  where  the spin current
$J$ and the von-Neumann entropy VNE, are depicted both as explicit
and as implicit functions of $\Delta$. From the top left panel of
Fig. \ref{Fig4} we see that, quite interestingly,  the system
undergoes current reversals in the critical region $-1 \le
\Delta\le 1$ (see the reversal occurring at $\Delta\approx 0.4$),
this increasing by two the number of peaks of $|J|$ discussed for
the  $N=3,4$ cases (notice that, due to the antisymmetry of
$J(\Delta)$ for $N$ odds, we restricted these figures to the range
$\Delta\ge 0$).
\begin{figure}[ptb]
\centerline{
\includegraphics[width=5.6 cm,height=7.6cm,clip]{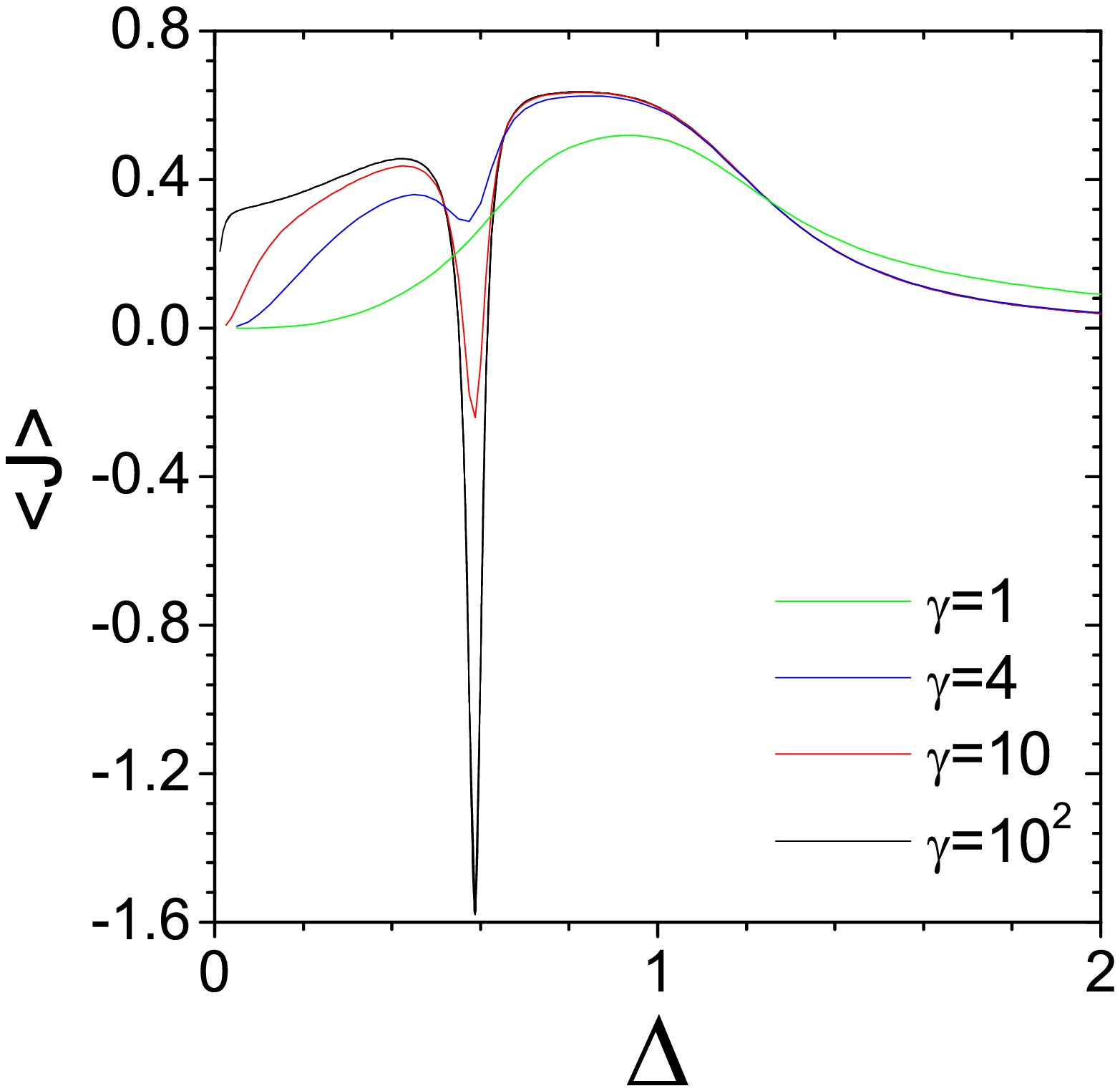}
\hskip -1.2cm
\includegraphics[width=5.6 cm,height=7.6cm,clip]{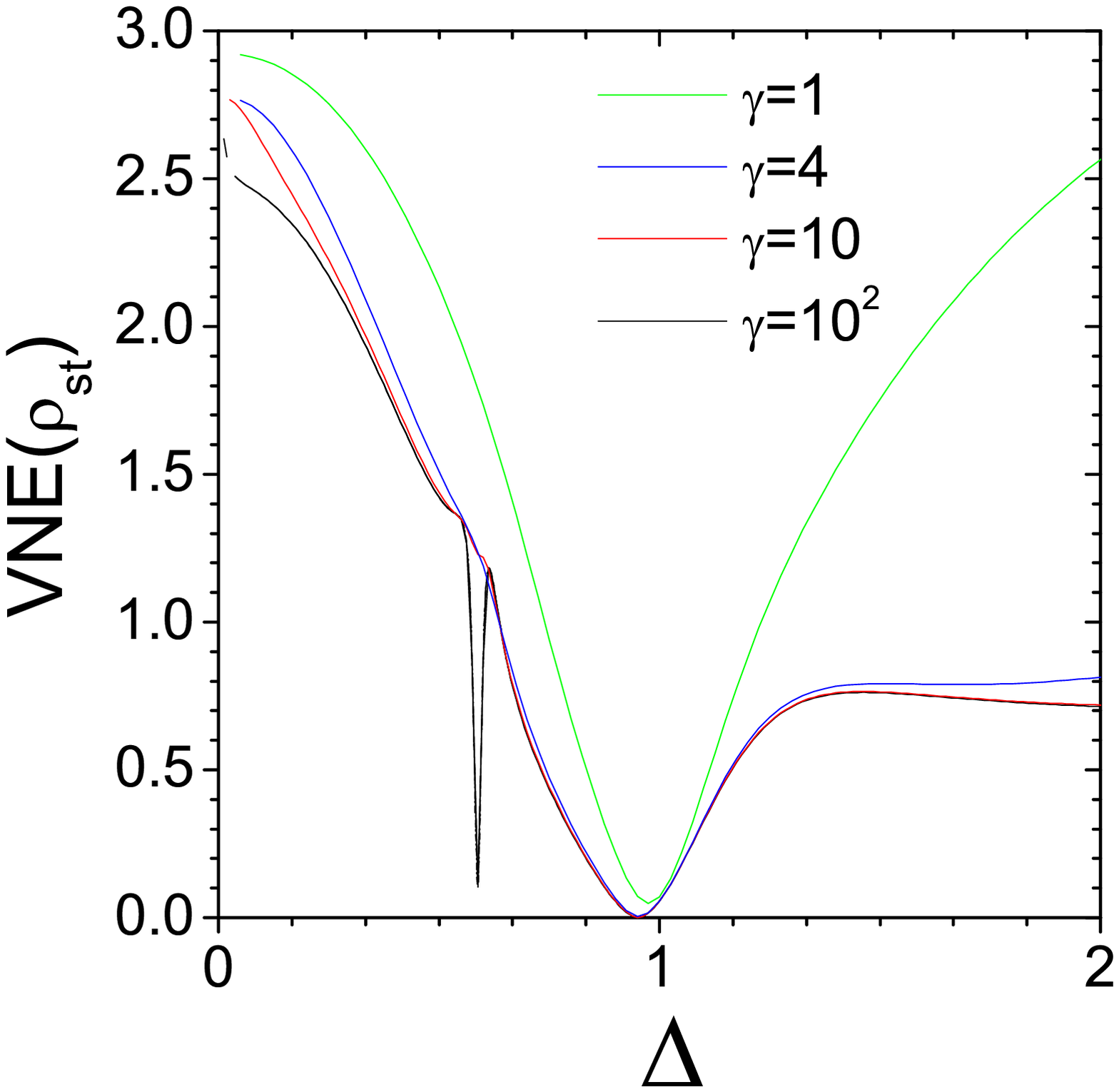}
}
\vskip -2cm
\caption{(Color online) Top panels.Stationary spin currents  $J$ (left panel) and von Neumann entropies (right panel)   vs $\Delta$ for the
$\mathcal{H}_{XXZ}$ chain of length $N=6$ for different values of $\gamma$. Curves are obtained from exact stationary solution of the Lindblad Master Equation. Due to the parity of $J(\Delta)$ for $N$ even, we have restricted the figure to the range $\Delta\ge 0$ only.
}
\label{Fig6}%
\end{figure}
The negative peaks in the current correspond to a relative minima
appearing in the VNE curve, as one can see from the top right
panel of this figure. Notice, however, that the correspondence
between  secondary peaks of $|J|$  and corresponding VNE minima
occurs with delay in $\gamma$ (notice that there  exist  tiny
minima in $J$ which have no corresponding minima in the VNE
curve). Details of the current reversal phase transition  and the
behavior of the corresponding  VNE curves  are  reported in the
bottom panels of  Fig. \ref{Fig4} for different values of the
coupling parameters. Also, in Fig. \ref{Fig5} we have depicted the
VNE vs stationary current $J$ as parametric plots for fixed
$\gamma$ and $\Delta$ varied in the range  $]-\infty, \infty[$.
\begin{figure}[ptb]
\centerline{
\includegraphics[width=8.6 cm,height=10.6cm,clip]{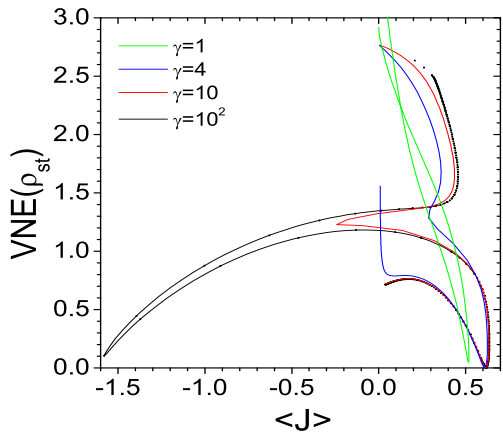}
} \vskip -2cm \caption{(Color online) von Neumann entropy VNE vs
stationary current $J$ for the open $\mathcal{H}_{XXZ}$ chain of
length $N=6$ for coupling constants $\gamma$ as indicated in the
panels.  As the coupling constant $\gamma$ is increased the values
of the VNE at the two minima decrease and in  the strong coupling
limit $\gamma \rightarrow \infty$ they reduce  exactly to zero
(pure states). }
\label{Fig7}%
\end{figure}
 From the panel insets one can see that the
minima of the VNE are always very close to  maxima of the
absolute value of the current. As the coupling constant $\gamma$ is increased
the values of the VNE at the minima decrease and in  the strong coupling limit
$\gamma \rightarrow \infty$ they reduce  exactly to zero (pure states), in analogy with  the cases $N=3,4$ discussed before.

Similar behavior, with appearance of reversal currents in the
critical region is observed also for the case $N=6$, as one can
see from Fig. \ref{Fig6}. The appearance  of the reversal current
can be seen as a precursor of  a quantum phase transition driven
by the
 boundaries as the coupling parameter is varied. As a general rule we conjecture  that
the number of extrema of $|J|$ is at most $N - 2$ times  for $N$
even and $N-1$ times for  N  odd. These extrema occur always in
the critical region and may disappear as the coupling constant is
decreased.

\section{Discussion and Conclusions}
We have investigated the transport properties of an  Heisenberg
XXZ chain in contact with twisted XY-boundary magnetic reservoirs
by means of the Lindblad master equation. Exact solutions for the
stationary density matrix have been constructed for chains of
small sizes using the quantum symmetry of the system. Using these
solutions we have  investigated the transport property  of the
chain   in terms of the von Neumann entropy. As a result we have
shown that the maximal spin current always occurs  in the
proximity of a minimum  of the VNE and for particular choices  of
the anisotropy parameters it can exactly coincide  with it. More
precisely, we found that as the coupling constant increases, the
VNE-$J$ curve undergoes a crossing point with the development of a
cusp for a particular value of $\Delta$ for which  the maximal
current is in exact coincidence with the minimum of the VNE. We
also showed the existence of current reversals e.g. the presence
of negative peaks in current, which correlate to relative minima
in the VNE curve, when the anisotropy parameter is  in the
critical region  $-1 \le \Delta\le 1$. These relative minima
disappear in the small coupling limit, while in the infinite
coupling limit we show that the minima of the VNE becomes exactly
zero, meaning that  maximal transport in this case is achieved
with states very close to pure states.

\vskip -4cm

\acknowledgements
MS acknowledges support from the
Ministero dell' Istruzione, dell' Universit\`{a} e della Ricerca
(MIUR) through a \textit{Programma di Ricerca Scientifica di
Rilevante Interesse Nazionale} (PRIN)-2010 initiative.

\vskip 0cm

\begin{widetext}
\appendix\section{\centerline{Exact stationary density matrix elements for the case $N=3$}}
 One can check that the most general form of the density matrix compatible with the symmetry operation $T$ in Eq. (\ref{trot}) for the case $N=3$, is
\begin{equation}
\rho=\left(
\begin{array}{cccccccc}
 a_{11} & a_{12}+i b_{12} & a_{13}+i b_{13} & a_{14}+i b_{14} & a_{15}+i b_{15} & a_{16}+i b_{16} & a_{17}+i b_{17} & (1-i) a_{18} \\
 a_{12}-i b_{12} & a_{22} & a_{23}+i b_{23} & (1+i) a_{24} & a_{25}+i b_{25} & a_{26}+i b_{26} & a_{27}+i b_{27} & i b_{14}-a_{14} \\
 a_{13}-i b_{13} & a_{23}-i b_{23} & a_{33} & i a_{26}+b_{26} & a_{35}+i b_{35} & (1+i) a_{36} & a_{37}+i b_{37} & i b_{16}-a_{16} \\
 a_{14}-i b_{14} & (1-i) a_{24} & b_{26}-i a_{26} & a_{22} & b_{27}-i a_{27} & a_{23}+i b_{23} & a_{25}+i b_{25} & i a_{12}+b_{12} \\
 a_{15}-i b_{15} & a_{25}-i b_{25} & a_{35}-i b_{35} & i a_{27}+b_{27} & a_{55} & i a_{37}+b_{37} & (1+i) a_{57} & i b_{17}-a_{17} \\
 a_{16}-i b_{16} & a_{26}-i b_{26} & (1-i) a_{36} & a_{23}-i b_{23} & b_{37}-i a_{37} & a_{33} & a_{35}+i b_{35} & i a_{13}+b_{13} \\
 a_{17}-i b_{17} & a_{27}-i b_{27} & a_{37}-i b_{37} & a_{25}-i b_{25} & (1-i) a_{57} & a_{35}-i b_{35} & a_{5
 5} & i a_{15}+b_{15} \\
 (1+i) a_{18} & -a_{14}-i b_{14} & -a_{16}-i b_{16} & b_{12}-i a_{12} & -a_{17}-i b_{17} & b_{13}-i a_{13} & b_{15}-i a_{15} & a_{11} \\
\end{array}
\right)
\end{equation}
 This can  be easily proved by substituting an arbitrary form of $\rho$, e.g. with arbitrary diagonal real elements, $a_{ii}$, and arbitrary off diagonal complex elements, $a_{ij}+ib_{ij}$,
 into Eq. (\ref{trota}) and by imposing the equality. This gives the set of relations reported in Eqs. (\ref{elem1}), (\ref{elem2}) and made explicit in the above $\rho$.

The exact expressions of $\rho_{ij}$ are obtained by substituting (A1) into the Lindblad master equation and by imposing the stationarity. This gives the following system of algebraic equations :
\begin{eqnarray}
&& 2 (a_{14}-a_{17})+2(b_{14}+b_{17})+a_{27}-b_{27}+6a_{18}=0, \nonumber \\
&& 2 a_{11}-a_{22}-a_{55}=0, \nonumber \\
&& \gamma (2a_{22}-a_{11}-a_{33})+4 b_{23}=0, \nonumber \\
&& \gamma (2a_{55}-a_{11}-a_{33})-4 b_{3,5}=0, \nonumber \\
&& \gamma (2a_{33}-a_{22}-a_{55})+4 (b_{35}-b_{23})=0, \nonumber \\
&& \gamma (2a_{24}-a_{13}-b_{13})+2 (b_{26}-a_{26})=0, \nonumber \\
&& \gamma (2a_{57}-a_{13}-b_{13})+2 (a_{37}-b_{37})=0, \nonumber \\
&& \gamma (a_{12}+b_{15}-b_{16}-3b_{25}+b_{26}+b_{37})+a_{23}-a_{35}=0, \nonumber \\
&& \gamma (a_{15}+a_{26}-a_{37}-3a_{16}-b_{12})-b_{14}-b_{17}+2\Delta b_{16}=0, \nonumber \\
&& \gamma (3a_{15}-a_{26})+2(b_{13}-\Delta b_{15})=0, \nonumber \\
&& \gamma (3b_{12}-a_{37})-2(a_{13}-\Delta a_{12})=0, \nonumber \\
&& \gamma (3a_{37}-b_{12})-2(a_{57}-a_{36}+b_{27}-\Delta b_{37})=0, \nonumber \\
&& \gamma (3a_{26}-a_{15})+2 (a_{24}-a_{36}+b_{27}-\Delta b_{26})=0, \nonumber \\
&& \gamma (3a_{25}-a_{15}-a_{26}-a_{37}-b_{12})-b_{35}+b_{23}=0, \nonumber \\
&& \gamma (2a_{13}-a_{24}-a_{57})+2(b_{12}+b_{15}-2\Delta b_{13})=0, \nonumber \\
&& \gamma (2b_{13}-a_{24}-a_{57})-2(a_{12}+a_{15}-2\Delta a_{13})=0,  \\
&& \gamma [2(a_{24}+a_{13})-4a_{14}-a_{17}-a_{23}]-2(b_{16}-\Delta b_{14})=0, \nonumber \\
&& \gamma [2(a_{24}+b_{13})-4b_{14}+b_{17}-b_{23}]+2(a_{16}-\Delta a_{14})=0, \nonumber \\
&& \gamma [2(a_{57}+a_{13})-4b_{17}+b_{14}-b_{35}]+2(a_{16}-\Delta a_{17})=0, \nonumber \\
&& \gamma [2(a_{57}+b_{13})+4a_{17}+a_{14}-a_{35}]+2(b_{16}-\Delta b_{17})=0, \nonumber \\
&& \gamma [2(a_{57}+b_{13})-4a_{35}+a_{23}+a_{17}]+2(b_{25}-\Delta b_{35})=0, \nonumber \\
&& \gamma [2(a_{57}+a_{13})-4b_{35}+b_{23}-b_{17}]+2(a_{33}-a_{55}-a_{25}+\Delta a_{35})=0,\nonumber \\
&& \gamma [2(a_{24}+a_{13})-4a_{23}-a_{14}+a_{35}]-2(b_{25}-\Delta b_{23})=0, \nonumber \\
&& \gamma [2(a_{24}+b_{13})-4b_{23}-b_{14}+b_{35}]+2(a_{22}-a_{33}+a_{25}-\Delta a_{23})=0, \nonumber \\
&& \gamma [2(a_{11}+a_{55})-5b_{15}+b_{26}]+2(a_{13}-\Delta a_{15})=0, \nonumber \\
&& \gamma [2(a_{11}+a_{22})-5a_{12}+b_{37}]-2 (b_{13} -\Delta b_{12})=0, \nonumber \\
&& \gamma [2(a_{33}+a_{22})-5b_{26}+b_{15}]+2(a_{24}+a_{27}-a_{36}-\Delta a_{26})=0, \nonumber \\
&& \gamma [2(a_{33}+a_{55})-5b_{37}+a_{12}]+2(a_{36}-a_{27}-a_{57}+\Delta a_{37})=0,\nonumber \\
&& \gamma [2(a_{14}-a_{17})+6a_{27}+a_{18}-a_{36}+2(b_{23}+b_{35})]+2(b_{26}-b_{37})=0, \nonumber \\
&& \gamma [2(a_{23}+a_{35})-6b_{27}+a_{18}-a_{36}+2(b_{14}+b_{17})]+2(a_{26}-a_{37})=0, \nonumber \\
&& \gamma [2(a_{23}+a_{35})-6a_{36}+a_{27}-b_{27}+2(b_{23}+b_{35})]-2(a_{26}-a_{37}-b_{26}+b_{37})=0, \nonumber \\
&& \gamma (a_{12}+b_{15}-3b_{16}-b_{25}+b_{26}+b_{37})+a_{14}+a_{17}-2\Delta a_{16}=0,\nonumber \\
\nonumber
\label{systemn3}
\end{eqnarray}
\end{widetext}
for the real and imaginary parts, $a_{ij}, b_{ij}$,  respectively,
of the density matrix elements. One can easily check that  the
solution becomes unique when the normalization condition
$Tr(\rho)=1$ is imposed, this completely solving the problem. This
approach, being general, can be used for other values of $N$.

In the limit $\gamma \rightarrow \infty$ the above set of equation
simplifies and yields a compact solution for the density matrix,
(see also \cite{slava}), in the form
\begin{equation}
\rho_{\gamma \rightarrow \infty}=\frac {1}{4} I + \frac {1}{4}
(I-\sigma^{y}) \otimes \frac{\Delta}{1+2\Delta^{2}}(\sigma^{x}
-\sigma^{y}) \otimes (I+\sigma^{x}). \label{rhoLimit_N3}
\end{equation}
From this expression we readily compute the  eigenvalues for the
reduced density matrix,
$\lambda_i=(0,0,0,0,0,0,\lambda_1,\lambda_2)$ where
\begin{equation}
\lambda_{1,2}=\frac{1+4 \Delta^2+4\Delta^4 \pm 2 \Delta \sqrt{2}
\sqrt{1+4 \Delta^2+4\Delta^4} } {2(1+4 \Delta^2+4\Delta^4)}.
\end{equation}
At the point $\Delta=1/\sqrt{2}$, $\lambda_1=0,\lambda_2=1 $ and
the density matrix spectrum becomes that of a pure state. The
point $\Delta=1/\sqrt{2}$ is exactly the  point where the steady
current attains its maximum value $max_{0\leq \Delta \leq \infty}
J(\Delta)=\sqrt{2}$.

Analogous calculations for higher number of sites $N>3$ lead to
similar results, with the difference that the steady state in the
limit of large couplings becomes pure not exactly at the point
$\Delta $ where the magnetization current has an extremum, but
quite close to it. For $N=4$, the extremum of the current at
$\gamma \rightarrow \infty $ reaches its maximum at the point
$\Delta =(-1/4+\sqrt{3}/2)^{1/2}\approx 0.785$ while the density
matrix spectrum becomes "pure" at $\Delta ^{\ast
}=\sqrt{3}/2\approx 0.866$. Note however that the maximum peak
$J(\Delta )$ is rather broad in this case, so that the maximal $j$
value\ and its value at  the "pure" point $J(\Delta ^{\ast })$ are
very close, $(J_{\max }-$ $J(\Delta ^{\ast }))/J_{\max }\approx
0.009$.   At $N=5$, there are two extrema, at $\Delta
_{1,2}\approx 0.384,0.814$, while the state becomes pure at the points $%
\Delta _{1,2}^{\ast }\approx 0.383,0.924$. In the point $\Delta
_{1}\approx 0.384$  the peak is narrow, and so $\Delta _{1}\approx
\Delta _{1}^{\ast }$, while the second peak at $\Delta _{2}\approx
0.814$ is broad one, justifying the discrepancy between the
$\Delta _{1}$ and $\Delta _{2}^{\ast }$ (data not shown).

\end{document}